\documentclass[fleqn,authoryear]{elsarticle}
\usepackage[british]{babel}
\usepackage[T1]{fontenc}
\usepackage{mathptmx}

\usepackage{amsmath}
\usepackage{amssymb}
\usepackage{url}
\usepackage{upgreek}
\usepackage{graphicx}
\usepackage{cleveref}
\usepackage{siunitx}
\usepackage{multirow}
\usepackage{enumitem}
\setitemize{noitemsep}

\renewcommand{\d}{\ensuremath{\text{d}}}

\begin{document}
  
\title{A new method to determine multi-angular reflectance factor from lightweight multispectral cameras with sky sensor in a target-less workflow applicable to UAV}
\author[pix4d]{Klaus Schneider-Zapp\corref{cor1}}
\ead{klaus.schneider-zapp@pix4d.com}
\author[pix4d]{Manuel Cubero-Castan}
\author[pix4d]{Dai Shi}
\author[pix4d]{Christoph Strecha}
\cortext[cor1]{Principal corresponding author}
\address[pix4d]{Pix4D, EPFL Innovation Park, Building F, 1015 Lausanne, Switzerland}

\begin{abstract}
A new physically based method to estimate hemispheric-directional reflectance factor (HDRF) from lightweight multispectral cameras that have a downwelling irradiance sensor is presented.
It combines radiometry with photogrammetric computer vision
to derive geometrically and radiometrically accurate data purely from the images, without requiring reflectance targets or any other additional information apart from the imagery.
The sky sensor orientation is initially computed using photogrammetric computer vision and revised with a non-linear regression comprising radiometric and photogrammetry-derived information.
It works for both clear sky and overcast conditions.
A ground-based test acquisition of a Spectralon target observed from different viewing directions and with different sun positions using a typical multispectral sensor configuration for clear sky and overcast showed that both the overall value and the directionality of the reflectance factor as reported in the literature were well retrieved.
An RMSE of \SI{3}{\percent} for clear sky and up to \SI{5}{\percent} for overcast sky was observed.
\end{abstract}


\maketitle

\section{Introduction}

Image-based remote sensing is a widely used tool, from Earth science disciplines like oceanography or geology to dedicated applications like mineral exploration or precision agriculture.
Traditionally, space-borne and air-borne sensors were used to retrieve reflectance factors in several spectral bands. Satellites have the advantage of large spacial coverage, e.g.\ to predict maize yield on the whole USA \citep{sakamoto+2014} or to monitor forest clearance in Madagascar \citep{page2002}.
However, they are limited by low resolution, long revisiting times, no available data under cloud cover, and high costs to operate and maintain.
Air-borne platforms allow better flexibility in sensors, naturally increasing spectral and spatial resolution, but are also too costly to allow short revisiting times \citep{matese+2015}. 

Recently, unmanned aerial vehicles (UAVs) have become popular sensing platforms.
They allow timely measurements, have high resolution and can be operated under cloud cover.
Light multi- and hyper-spectral sensors which are suitable to be flown on UAVs have become available in the last years \citep{colomina+2014b}.
Especially in precision agriculture, where the temporal evolution needs to be tracked, UAVs have become an important monitoring tool, which is used to estimate plant properties like leaf area index \citep{lelong+2008} or biomass \citep{bendig+2015}, to detect diseases \citep{nebiker+2016}, or to predict yields \citep{geipel+2014,nebiker+2016}.

Radiometric accuracy is of prime importance for accurate results. Artefacts due to incorrect radiometry can lead to misinterpretation.
Traditional methods from satellites are not readily applicable to UAV-based data due to different spacial scales, illumination conditions (satellites only yield data without cloud cover), and instrument constraints.
To date, most researchers use the empirical linear method to calibrate their sensor with known reflectance targets \citep{jakob+2017,wang+2015}.
\citet{hakala2010} derived the reflectance factor by using a known reflectance target which must be visible in every single image; this is impractical for any larger area or frequent measurements as required by precision agriculture monitoring for instance.
\citet{honkavaara+2018} use an optimisation procedure to derive radiometric parameters from reflectance targets.
However, as illumination conditions can change rapidly under cloud cover, its applicability is quite limited, since the reflectance target has to be measured repeatedly for every change in illumination.
Furthermore, for practical data acquisition of larger areas, a simple workflow is required.
This is especially true for applications like precision agriculture, where operators may not be experts and it is important to reduce any human intervention as much as possible.

Multispectral cameras with a ``sky sensor'' that captures the downwelling irradiance have recently become commercially available.
With the correct camera calibration, those sensors can be used to derive reflectance factors without the need of an empirical calibration procedure for every acquisition \citep{pix4d_wispers2018}.
Traditional correction methods like \citet{li+2016,colby1991} which are based on work for satellites or airborne measurements only work for clear sky, as they neglect the contribution of scattered skylight.
The International Commission on Illumination (Commission Internationale de l'Eclairage, CIE) published a standard sky model \citep{darula+2002} that describes the sky illuminance for a broad range of sky conditions, from overcast to clear sky.
It has recently been used for estimating the illuminance on tilted areas \citep{sola+2016}.
However, due to the large number of parameters, it is not easily usable for UAV data.
Especially for overcast sky, where the illuminance can be a quickly changing function of time, an estimation only from the images is problematic.

The desired outcome for most multispectral acquisitions are generally reflectance maps, or in some case the computation of vegetation indices based on those reflectance maps.
By definition, reflectance is a multi-angular quantity \citep{schaepmanstrub+2006} and reflectance maps are an estimation of these quantities at a given viewing angle (normally nadir).
Methods have been designed to correct for this angular effect.
Such corrections are specifically required for time series analysis \citep{roujean+1992}.
Models based on multi-angular acquisition are used to recover different parameters, like the canopy structure of forests \citep{widlowski+2004} or advanced vegetation indexes like Leaf Area Index or foliage clumping index \citep{chen+2005, schaepman+2007}.
An advantage of UAV acquisitions is that measurements are multi-angular: a good overlap between different frames is usually required, resulting in measuring all scene materials with different viewing angles for each acquisition.
This allows to perform an analysis similar to what was previously only possible with satellite data while still using all advantages of UAVs.

In this paper, we describe a new physically based method to derive the hemispherical-directional reflectance factor (HDRF) with a multispectral camera and a sky sensor.
The proposed method works in both clear sky and overcast conditions by describing both the direct and scattered sunlight contributions to the illumination, and takes into account the local surface orientation as well as the sky sensor orientation.
All required parameters are estimated solely from the multispectral images by combining photogrammetry and radiometry.

First we quickly review radiometric corrections for image acquisition.
Then we describe our new algorithm to take the sky sensor data into account for estimating HDRF in a target-less workflow.
We present the data acquisition and processing of the verification experiment.
Finally we discuss the results and show that our new method is able to correctly retrieve HDRF in both overcast and clear sky conditions.

\section{Review of radiometry of imaging}
\label{sec:review}

The basics of radiometry and remote sensing are treated comprehensively in textbooks, e.g.\ \citet{mccluney2014,jaehne2012,schott2007}. An overview of reflectance quantities is found in \citet{schaepmanstrub+2006}. Here we only quickly review the most important concepts required for this study. Only the optical and near-infrared spectrum are considered. In this spectral range, the observed signal is dominated by reflected sunlight.

Consider a sun-illuminated object on a field observed by a multispectral camera with a sky sensor aboard a UAV. 
A typical height is around \SI{100}{m}.
The object receives direct solar irradiance (directional), and scattered radiance (skylight and light scattered in clouds) from the entire hemisphere \citep{schott2007}.
If there are no high objects in the field, light reflected by adjacent objects and multiple scattering can be neglected.
Due to the low flying height, atmospheric scattering on the path between the object of interest and the camera (light from the object that is scattered away from the camera and light scattered in the atmosphere into the camera) can be neglected and the down-welling irradiance reaching the UAV is practically identical to the one on the ground \citep{schlapfer+2005}.
The instantaneous field of view of one pixel of a typical multispectral camera is very small (e.g.\ for the Parrot Sequoia it is \SI{0.055}{\degree} horizontally and vertically), thus the solid angle is considered infinitesimal.
The quantity of interest is the hemispherical-directional reflectance factor (HDRF), which is defined as
\begin{equation}
\label{eq:hdrf-def}
R(\theta_\text{i}, \varphi_\text{i}, 2\pi; \theta_\text{r}, \varphi_\text{r})
 = \frac{\int\limits_{2\pi} f_\text{r}(\theta_\text{i}, \varphi_\text{i}, \theta_\text{r}, \varphi_\text{r}) L_\text{i}(\theta_\text{i}, \varphi_\text{i}) \, \d\Omega_\text{i}}
   {(1/\pi) \int\limits_{2\pi} L_\text{i}(\theta_\text{i}, \varphi_\text{i}) \, \d\Omega_\text{i}}
   \,,
\end{equation}
where
\begin{equation}
f_\text{r} = \frac{\d L_\text{r}}{\d E_\text{i}}
\end{equation}
is the bi-directional reflectance distribution function (BRDF),  $L_\text{i}$ the incoming spectral radiance, $E_\text{i} = \int\limits_{2\pi} L_\text{i} \, \d\Omega_\text{i}$ the incoming spectral irradiance,  $L_\text{r}$ the reflected spectral radiance, $\theta_\text{i}$ and $\varphi_\text{i}$ the zenith and azimuth angles of the incoming direct sunlight, $\theta_\text{r}$ and $\varphi_\text{r}$ the zenith and azimuth angles of observation, and $\Omega$ solid angle \citep{schaepmanstrub+2006}.

We divide the incoming radiance $L_\text{i}$ into direct sunlight $E_\text{d}$ with angles $\theta_\text{i}$ and $\varphi_\text{i}$ (through a hypothetical surface perpendicular to the beam direction) and diffuse scattered sunlight $L_\text{s}$.
Assume that the diffuse part is isotropic (i.e.\ independent of the angles).
Then $L_\text{s}$ is a constant and the diffuse irradiance received from the whole hemisphere is $E_\text{s} = \int\limits_{2\pi} L_\text{s} \, \d\Omega = \pi L_\text{s}$.
Let
\begin{equation}
  \label{eq:direct-sunlight-ratio-def}
  \epsilon = \frac{E_\text{d}}{E_\text{d} + \pi L_\text{s}}
\end{equation}
be the direct sunlight ratio, i.e.\ the ratio of direct irradiance to total irradiance.
With this definition, we can re-write \citet{schaepmanstrub+2006}, eq.~(12)--(13), as
\begin{equation}
  \label{eq:hdrf-from-brdf}
  R(\theta_\text{i}, \varphi_\text{i}, 2\pi; \theta_\text{r}, \varphi_\text{r})
  = \epsilon \pi f_\text{r}(\theta_\text{i}, \varphi_\text{i}, \theta_\text{r}, \varphi_\text{r}) + 
  (1 - \epsilon) \int\limits_{2\pi} f_\text{r}(\theta_\text{i}, \varphi_\text{i}, \theta_\text{r}, \varphi_\text{r}) \, \d\Omega_\text{i}
  \,.
\end{equation}
For a Lambertian reflector, the equation simplifies to $R = \pi f_\text{r}$.

A pixel of an optical camera with focal length $f$ and aperture number $k = f/a$ at distance $d$ from a Lambertian source of spectral radiance $L_\uplambda$ receives the spectral irradiance
\begin{equation}
\label{eq:radio-model-jaehne}
E'_\uplambda = t \frac{\pi \cos^4 \theta}{4 k^2 (1 + m)^2} L_\uplambda
\,,
\end{equation}
where $\theta$ is the observation angle, $m = \frac{x'}{x} = \frac{f + d'}{f + d} = \frac{f}{d}$ the projection scale, and $t$ the transmissivity of the optical system (a function of wavelength and observation angle) \citep[Sect.~3.5]{jaehne2012}. A solid state imaging sensor such as a complementary metal-oxide-semiconductor (CMOS) sensor can be modelled as
\begin{equation}
\label{eq:sensor_p}
p = p_0 + K \tau A_\text{pix} \int\limits_0^\infty \eta E'_\uplambda \d\lambda
\,,
\end{equation}
where $p$ is the pixel digital number (DN), $p_0$ the dark current (signal obtained without any radiance, caused by thermal stimulation), $K$ the gain, $\tau$ the exposure time, $A_\text{pix}$ the area of the pixel, $\eta$ the quantum yield, and $\lambda$ wavelength \citep[Sect.~6.4]{jaehne2012}.
Real sensors may have additional effects.
Electron leakage from the pixel well can result in a loss of signal.
For CMOS sensors with an electronic global shutter, light leakage into the hold cell can lead to an additional signal.

Introducing the effective radiance for the band of interest
\begin{equation}
L = \int\limits_0^\infty S L_\uplambda \, \d\lambda
\end{equation}
with the spectral sensitivity of the sensor $S$ (which is composed of the spectral variance of both the transmissivity $t$ and the quantum yield $\eta$) we can combine \cref{eq:radio-model-jaehne} and \cref{eq:sensor_p} as
\begin{equation}
\label{eq:p(L)}
p = p_0 + \frac{\tau K s v}{k^2} \frac{1}{(1 + m)^2} \pi L
\,,
\end{equation}
where $s$ is the (wavelength-independent) sensitivity and $v$ vignetting (composed of $\cos^4 \theta$ and the angle-dependent part of the transmissivity $t$).

\section{HDRF estimation from multispectral camera with sky sensor}
\label{sec:sun-model}

\begin{figure}
    \centering
    \includegraphics[width=.6\linewidth]{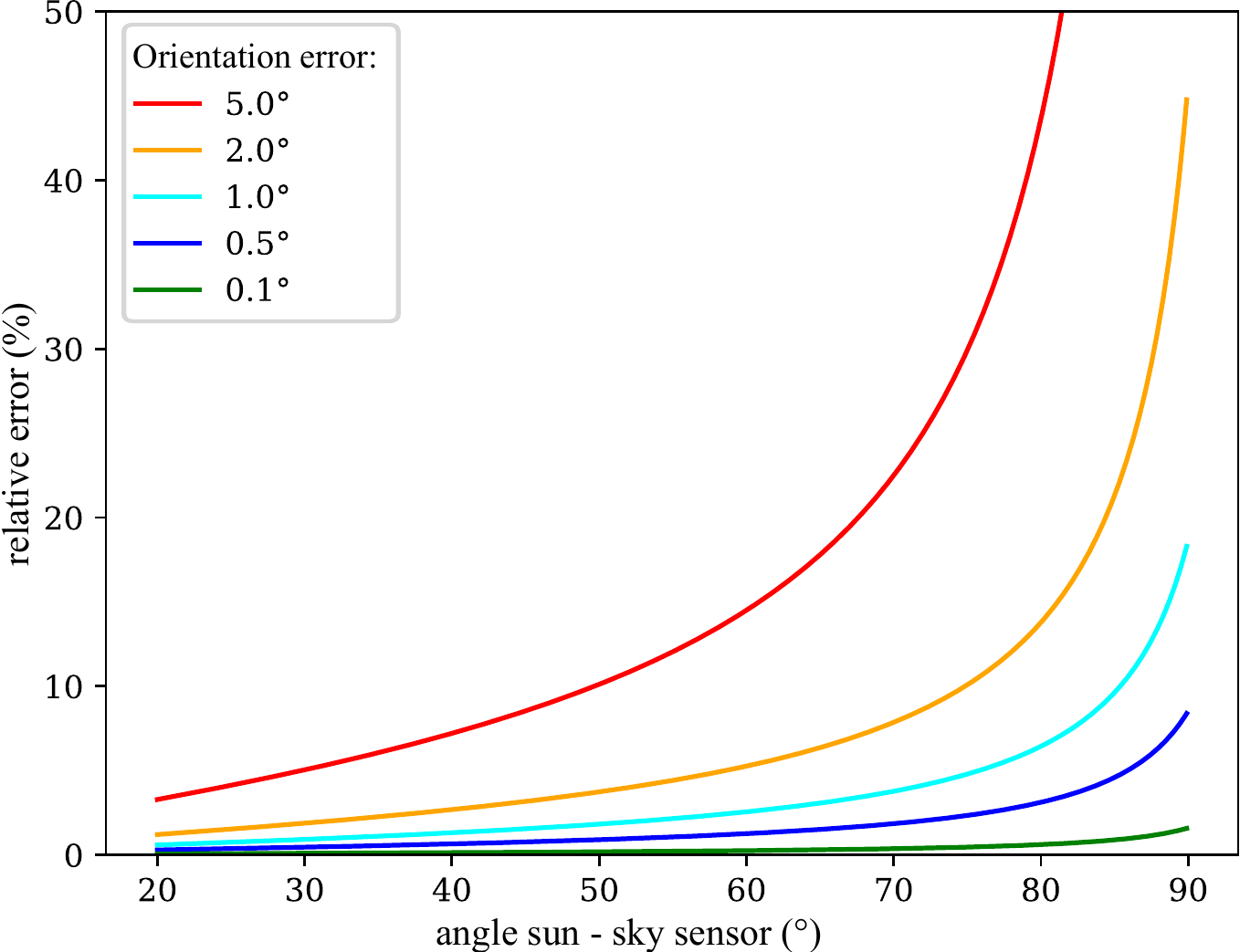}
    \caption{%
        Theoretical error in reflectance due to an incorrect sky sensor orientation as a function of the angle between sky sensor and sun for an orientation error of \SI{5.0}{\degree} (red), \SI{2.0}{\degree} (orange), \SI{1.0}{\degree} (cyan), \SI{0.5}{\degree} (dark blue), and \SI{0.1}{\degree} (green). The direct sunlight ratio is $\epsilon = 0.9$ and the sun zenith angle is $\SI{39.1}{\degree}$.}
    \label{fig:sky-sensor-orientation-error}
\end{figure}

This section describes our model to estimate HDRF by combining radiometry with photogrammetric computer vision.
It is applicable to any calibrated multispectral camera with sky sensor where the sky sensor is designed such that the orientation of the light-sensitive area is well-defined and the sensitivity does not change with angle, and which is rigidly connected to the camera body.
Initially, the scene is reconstructed with photogrammetric computer vision.
This results, among others, in the camera orientation and position in full 3D for each capture, and a geo-referenced digital surface model of the scene which includes the elevation and local surface normal of each point.
The lens distortion is also optimised during this process and automatically corrected when projecting the images.
It is assumed that the different bands of the multispectral camera and the sky sensor are mechanically rigidly connected and the captures of the different bands and the sky sensor are synchronised.
This allows to use the known constraints between the cameras for the different bands and the sky sensor to improve the speed and geometric accuracy.
We assume that the sky sensor measures the sun radiation at the same spectral band as the camera.
It is important that the orientation of the sky sensor is precise, since small errors in the angle between the sky sensor and the sun ray direction lead to significant errors in reflectance, especially when the angle between the sky sensor normal and the sun ray is large in clear sky conditions.
This is illustrated by \cref{fig:sky-sensor-orientation-error} which shows the theoretical relative error in reflectance caused by an error in the angle between sun and sky sensor due to an incorrect sky sensor orientation.
A major novelty of our method is the precise estimation of the sky sensor pose due to the combination of photogrammetry and radiometry, which is a prerequisite for obtaining reliable reflectance values.

\begin{figure}
	\centering
	\includegraphics[]{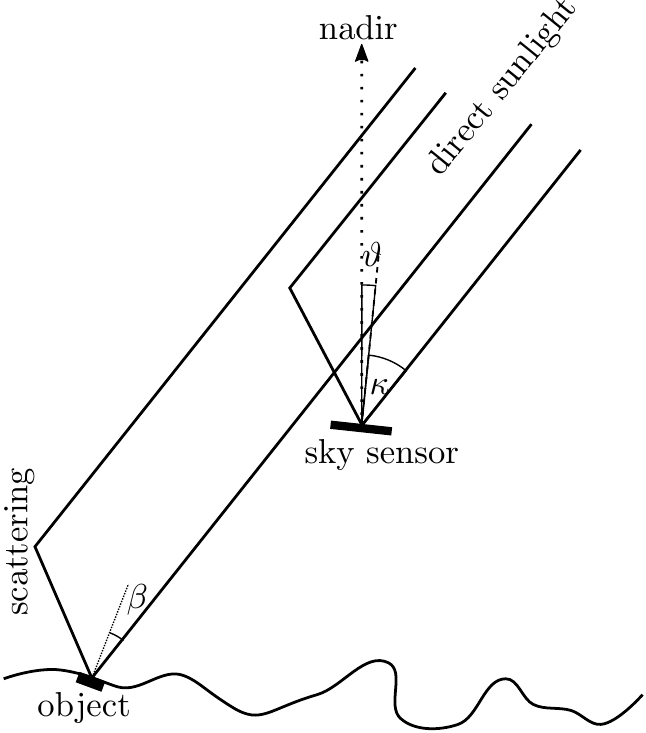}
	\quad
	\includegraphics[]{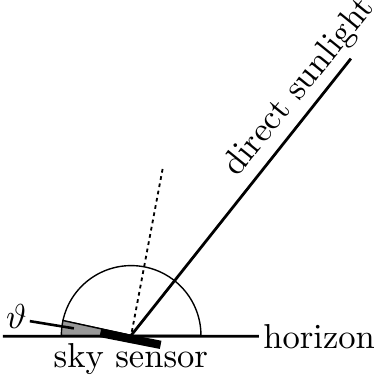}
	\caption{Geometry of sun radiation on the object and on the sky sensor.}
	\label{fig:sun-sensor}
\end{figure}

As discussed above, the sunlight consists of direct sunlight $E_\text{d}$ (referring to an area perpendicular to the sun direction) and diffuse scattered light $L_\text{s}$ (\cref{fig:sun-sensor}) and we assume that the scattered light is isotropic.
Let $\kappa$ be the angle between the normal of the sky sensor and the direct sunlight beam.
This angle is computed from the photogrammetrically derived sky sensor pose and the sun ray direction.
If the sky sensor is perfectly horizontal ($\vartheta = 0$) and measures the full hemisphere, the observed irradiance is
\begin{equation}
E_\text{sun,meas} = E_\text{d} \cos\kappa + L_\text{s} \int\limits_{2\pi} \d\Omega = E_\text{d} \cos\kappa + E_\text{s}
\,.
\end{equation}
In case $\kappa > \pi/2$, the sky sensor does not see any direct sunlight any more and the term $E_\text{d} \cos\kappa$ should be forced to zero.
If the sky sensor is inclined with respect to the horizon, $\vartheta \neq 0$, then even a sensor with perfect hemispherical sensitivity will not capture the complete scattered irradiance; the energy outside the field of view (the grey area in \cref{fig:sun-sensor} (right)) is missing:
\begin{equation}
\label{eq:sun-irradiance-meas}
E_\text{sun,meas} = E_\text{d} \cos\kappa + L_\text{s} \Omega_\text{v} = E_\text{d} \cos\kappa + \frac{\Omega_\text{v}}{\pi} E_\text{s}
\end{equation}
with the projected solid angle of view
\begin{equation}
\Omega_\text{v} = 
\begin{cases}
\int\limits_{\varphi=0}^\pi \int\limits_{\theta=0}^{\pi/2 - \vartheta} t(\theta, \varphi) \cos\theta \sin\theta \, \d\theta \d\varphi 
+ \int\limits_{\varphi=\pi}^{2\pi} \int\limits_{\theta=0}^{\pi/2} t(\theta, \varphi) \cos\theta \sin\theta \, \d\theta \d\varphi
&, 0 \leq \vartheta \leq \pi/2
\\
\int\limits_{\varphi=0}^\pi \int\limits_{\theta=\vartheta - \pi/2}^{\pi/2} t(\theta, \varphi) \cos\theta \sin\theta \, \d\theta \d\varphi 
&, \pi/2 < \vartheta \leq \pi
\end{cases}
\,,
\end{equation}
where $t_\text{s}(\theta, \varphi)$ is the directional transmissivity of the sky sensor. Note that the integral is written in the coordinate system of the sky sensor; the term $\cos\theta$ describes the projection of the incoming radiance onto the sky sensor. Only the fraction $\zeta = \frac{\Omega_\text{v}}{\pi}$ of the real scattered sun irradiance is measured. For an ideal sky sensor with constant transmissivity $t_\text{s} = 1$ over the whole hemisphere, we can evaluate the integral to $\Omega_\text{v} = \pi \left( \frac{1}{2} + \frac{1}{2}\cos^2\theta \right)$ for $0 \leq \vartheta \leq \pi/2$ and $\Omega_\text{v} = \pi \left( \frac{1}{2} - \frac{1}{2}\cos^2\theta \right)$ for $\pi/2 < \vartheta \leq \pi$.

Using the direct sunlight ratio $\epsilon$, we can compute the total irradiance received by a hypothetical area perpendicular to the sun direction:
\begin{equation}
\label{eq:sun-total-irradiance}
E_\text{sun,tot} = E_\text{d} + E_\text{s} = \frac{E_\text{sun,meas}}{\epsilon \cos\kappa + \zeta (1 - \epsilon)}
\,.
\end{equation}
If we know the direct light ratio $\epsilon$, we can estimate the sun irradiation on the object of interest:
\begin{equation}
\label{eq:obj-sun-irradiance}
E_\text{sun} = E_\text{sun,meas}\frac{\epsilon \cos\beta + (1 - \epsilon)}{\epsilon \cos\kappa + \zeta (1 - \epsilon)}
\,.
\end{equation}
As mentioned before, the local surface normal of each point is known from the photogrammetric reconstruction, therefore $\beta$ can be readily computed without any additional information.

Analogous to a camera pixel, the sky sensor measures a DN
\begin{equation}
\label{eq:irradiance-sensor-meas}
I = I_0 + \tau_\text{s} K_\text{s} s_\text{s} 
\int\limits_{0}^\infty S E_\text{sun,meas} \d\lambda
\,,
\end{equation}
where $I_0$ is the dark current, $K_\text{s}$ the gain, $\tau_\text{s}$ the exposure time, and $s_\text{s}$ the sensitivity of the sky sensor (the sky sensor is like a camera with only one pixel with an IFOV of the whole hemisphere).
The spectral sensitivity $S$ is assumed to be the same as the one of corresponding camera.

In a local coordinate system with the axes east, north, up, the direction from the sky sensor towards the sun is $\vec d_\text{sun} = (\sin\varphi_\text{i}, \cos\varphi_\text{i}, \cot\theta_\text{i})$.
As mentioned before, the direction of the sky sensor $\vec d_\text{sensor}$ is known from the photogrammetric reconstruction.
Then the cosine of the angle between the sun and the sensor is computed from the two vectors.

If the reflectance does not vary too much with the incident angles, we can approximate the integral in \cref{eq:hdrf-from-brdf} and the equation simplifies to $R = \pi f_\text{r} = \pi L_\text{r} / E_\text{i}$. 
By solving \cref{eq:p(L)} for $L$ and inserting that and \cref{eq:obj-sun-irradiance}, and using \cref{eq:irradiance-sensor-meas} to compute DN for the sunshine sensor, we can write the HDRF as a function of the measured quantities:
\begin{equation}
R = 
 \frac{(p - p_0) k^2 (1 + m)^2}{\tau K s v} \,
 \frac{\tau_\text{s} K_\text{s} s_\text{s}}{I-I_0} \,
 \frac{\epsilon \cos\kappa + \zeta (1 - \epsilon)}
 {\epsilon \cos\beta + (1 - \epsilon)}
 \,.
\end{equation}

\medskip

The direct sunlight ratio is not known a priori, thus it must be estimated.
Our method derives it directly from the measurements without requiring additional know\-ledge by fitting the angle-dependence of the sky sensor measurements. 
If there are no scattered clouds with sunny spots (e.g.\, scattered cumulus), the total downwelling irradiance $E_\text{sun,tot}$ is a smooth function of time, and for sufficiently small times it can be parametrised in a simple way.
We can use measurements of the irradiance $E_{\text{sun,meas}, i}$ and angles $\kappa_i$ to least-square fit $\epsilon$ and the parameters of the parametrisation:
\begin{equation}
\label{eq:sun-condition-fit}
\chi^2 = \frac{1}{N^2} \sum\limits_{i = 0}^N \left( E_\text{sun,tot} - \frac{E_{\text{sun,meas},i}}{\epsilon \cos\kappa_i + \zeta (1 - \epsilon)} \right)^2 \to \min
\,.
\end{equation}

The nominal relative angle between camera and sky sensor typically has an uncertainty of a few degrees, because the lenses always have a slight misalignment due to manufacturing tolerances, temperature changes, or shocks on the camera.
Therefore, we coupled our optimisation with the photogrammetric computation to be able to optimise the relative angle between the camera and the sky sensor.
The total downwelling irradiance was modelled as a polynomial.
The result of the fit is the direct sunlight ratio $\epsilon$, the total downwelling irradiance, and optionally a revised estimate of the relative angle.

\section{Data acquisition and processing}
\label{sec:method}

To test our new HDRF retrieval method, we conducted test experiments by acquiring images of a target of known reflectance under different viewing angles and sun to sky sensor angles, in order to compare the retrieved HDRF with its literature values.
The ground experiment avoids additional uncertainties that would be introduced by UAVs and allows more rigorous testing of our new method.

\begin{figure}
    \centering
    \includegraphics[width=.3\textwidth]{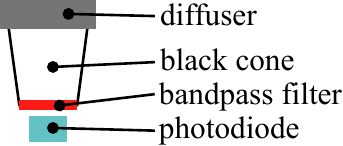}
    \caption{%
        Schematic of one band of the Sequoia sky sensor. A diffuser diffuses the incoming light. The bandpass filter is identical to the one used for the CMOS sensor. The setup is repeated for each band in a $2\times 2$ grid.}
    \label{fig:sky-sensor-schematic}
\end{figure}

The acquisitions were performed with the Parrot Sequoia multispectral camera\footnote{https://www.parrot.com/business-solutions-us/parrot-professional/parrot-sequoia/}, which is composed of 4 CMOS sensors with different narrow-band filters in the visible and near infrared domain; the spectral characterisation is detailed in \cref{tab:sq_bands}.
The camera comes with a sky sensor (Parrot call it ``sunshine sensor'') which measures the downwelling irradiance at the same spectral bands as the cameras.
Each band has its own sensing setup which is sketched in \cref{fig:sky-sensor-schematic}.
The four bands are arranged in a $2\times 2$ grid in the same way as the cameras.
Each band is composed of a diffuser, a spectral filter and a photo diode.
The diffuser diffuses the incoming light in order to provide a well-defined area for the projection of the incoming ray and to avoid any change in sensitivity of the photo diode with incoming ray angle.
The bandpass filter is identical to the one used for the camera and placed above the photodiode.
The spectral width of the bandpass filter is small compared to the spectral change in sensitivity of the photodiode and of a pixel of the CMOS sensor, therefore the spectral response shape of the sky sensor is virtually identical to the one of the camera.

In order to be able to retrieve reflectance without the need for a reflectance target, each camera unit is calibrated in the factory.
This includes independent measurements of each band for both sky sensor and camera.
\citet{pix4d_wispers2018} have shown a ground-based verification measurement to assess the precision and accuracy of this camera/sky sensor combination.
It involved a calibration target composed of 42 different grey patches (6 rows by 7 columns) whose true reflectance factors were measured using a spectro-radiometer. An accuracy of \SI{2.5}{\%} for all 4 bands was observed.

\begin{table}
\caption{Details of sequoia bands.}
\centering
\begin{tabular}{ l  l  l }
\hline
Band name & Wavelength (nm) & FWHM (nm) \\
\hline
Green & 550 & 40 \\
Red & 660 & 40 \\
Red edge & 735 & 10 \\
NIR & 790 & 40 \\
\hline
\end{tabular}
\label{tab:sq_bands}
\end{table}

Since the red edge band overlaps with a water absorption band, we computed the absorption due to water vapour using HITRAN data \citep{HITRAN2016}, assuming a homogeneous layer of \SI{100}{m} thickness (a typical flying height) with \SI{100}{\%} relative humidity at \SI{300}{K} and \SI{1013}{hPa} as a bad-case scenario, and a perfectly Gaussian sensitivity distribution with the centre and FWHM of each band.
The resulting absorbances are \SI{0.02}{\%}, \SI{0.1}{\%}, \SI{1.1}{\%}, and \SI{0.5}{\%}, for the green, red, red edge, and NIR bands, respectively.
Even the absorption for the red edge band of just above \SI{1}{\%} is below the measurement accuracy of the camera.

As reflectance target, we used a \SI{10}{cm} square white Spectralon\textsuperscript{\textregistered{}}, which is a sintered polytetrafluoroethylene (PTFE) material manufactured by Labsphere and widely used as standard target.
Its reflectance, close to 0.99, can be traced back to a NIST standard.
\citet{bruegge+2001} have acquired a BRF measurement of Spectralon and made it available online.

\begin{figure}
	\centering
	\includegraphics[width=.5\textwidth]{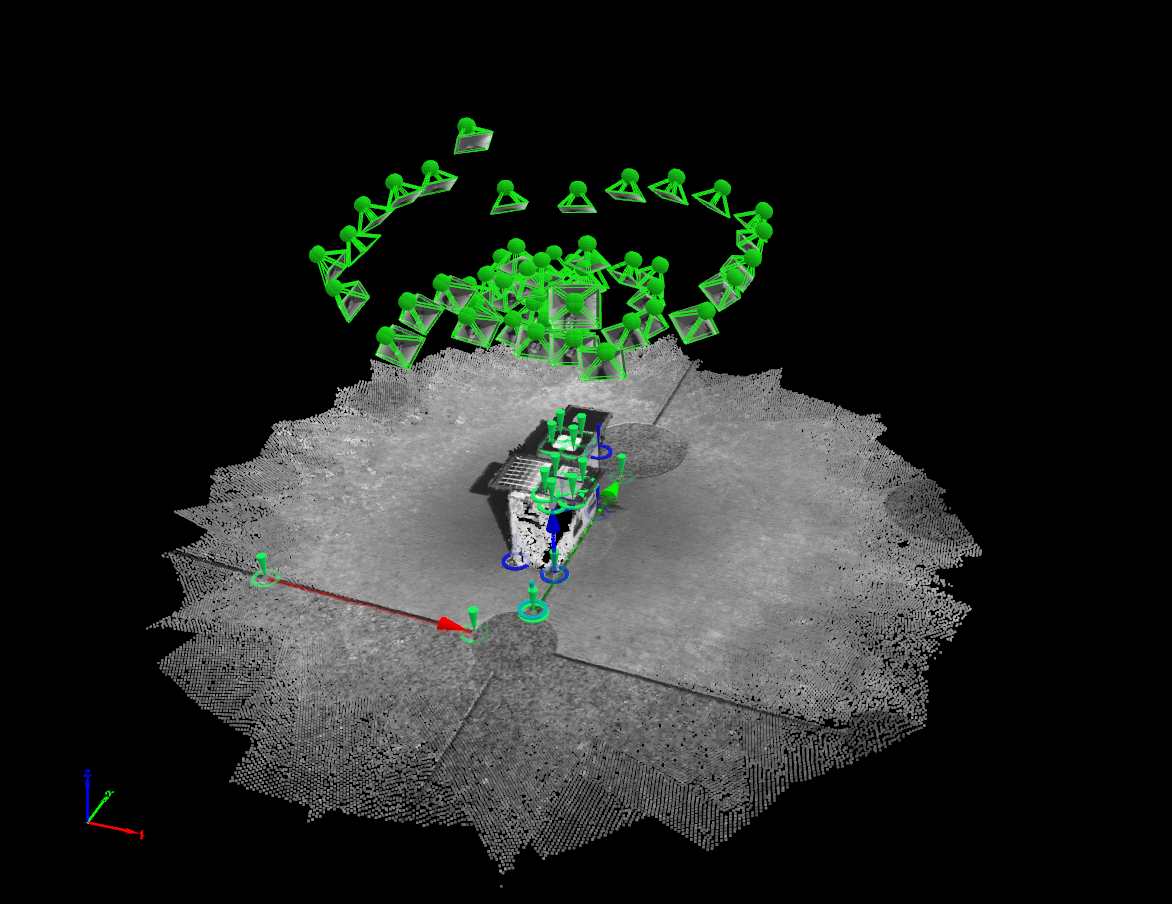}
	\caption{Rendering of the 3D model of one acquisition taken in clear sky condition as produced by Pix4Dmapper. In the centre is the box with the spectralon target. The position and orientation of the image captures is shown as green pyramids. GCPs are shown as green nails, the green, red, and blue arrows show the orientation of the North, East, and Up axis, respectively, as given by connecting appropriate GCPs.}
	\label{fig:hdrf_setup}
\end{figure}

The acquisitions have been performed under different light conditions (clear sky and overcast sky), at an open space at \'Ecole Polytechnique F\'ed\'erale de Lausanne (EPFL), Lausanne, Switzerland (\ang{46;31;06.2}\,N, \ang{6;33;59.4}\,E), which is flat and without any 3D structures close by, in order to prevent reflected light from adjacent objects.
The synthetic scene was comprised of a Spectralon target fixed on top of a box ($\SI{65}{cm}\times\SI{45}{cm}\times\SI{25}{cm}$).
The surrounding was a planar concrete place.
The sky sensor was mechanically rigidly fixed to the camera, with a known relative angle of $(\SI{175}{\degree}, \SI{0}{\degree}, \SI{0}{\degree})$ in $(x, y, z)$ directions of the camera reference (i.e.\ nearly opposite).
This is the configuration used for the Sequoia aboard the senseFly eBee, a widely used UAV.
The camera/sky sensor combination was fixed on a stick to avoid shadows by the operator.
Several Ground Control Points (GCPs) were acquired with an RTK (Real Time Kinematic) GPS (Javad, Triumph - LS, accuracy \SI{3}{cm} absolute) to georeference the scene.
A precise georeferencing is required to compute the correct relative angle between the sun and the sky sensor.
Images were taken at different angles, all around the box, in order to cover well the space of possible sun to sky sensor angles and observation angles.
\Cref{fig:hdrf_setup} shows the 3D scene reconstruction of the setup for one of the measurements, together with a visualisation of the position and orientation of the acquired images.
This also visualises the different observation angles used.

With clear sky, the irradiance is dominated by direct sunlight.
Hence, the computed HDRF can be compared to the laboratory-derived Spectralon BRF measurement by \citet{bruegge+2001} with the irradiance incident angle equal to the sun zenith angle.
Those database measurements were made with a Laser in a laboratory setting.
Originally, the data were used for space-borne measurements, which comprise only direct sunlight since there is no atmosphere.
This differs from our setup with clear sky, which also comprises a small diffuse part.
That diffuse component will make a small additional uniform contribution to the reflectance but not disturb the angular response.
Thus the error made by neglecting the diffuse part for estimating the directionality is small.
Measuring the true values in-situ was not feasible, because it is very costly and difficult to conduct.
The database contains the BRF of Spectralon measured with illumination angles of \SI{8}{\degree}, \SI{40}{\degree}, \SI{45}{\degree}, \SI{50}{\degree}, and \SI{55}{\degree}.
In order to compare to those values, we needed to perform our measurements at sun zenith angles covered by the database.

The overcast condition is dominated by diffuse light, therefore for the comparison, one needs to integrate over all incident angles.
The available data from \citet{bruegge+2001} is very sparse, only comprising five zenith angles.
Since all values are constant within error, we approximated it with a constant function in order to do the integration.

Clear sky acquisitions were performed on 21 September 2017, at 11am, 12pm, 3pm, and 4pm local time, with zenith angles of \SI{53}{\degree}, \SI{49}{\degree}, \SI{51}{\degree}, and \SI{55}{\degree}, respectively, to match the laboratory measurements for \SI{50}{\degree} and \SI{55}{\degree}.
Overcast sky acquisitions were performed on the 11 January 2018 at 9:30am local time.

Photogrammetric and radiometric image processing was done with a development version of Pix4Dmapper 4.3\footnote{\url{https://pix4d.com/product/pix4dmapper-photogrammetry-software/}}.
Image orientation was computed with advanced aerial triangulation (AAT) and bundle block adjustment (BBA), the project was georeferenced, a point cloud was generated, and the images were radiometrically corrected to HDRF.
The sky sensor orientation was computed from the photogrammetrically derived camera orientation using known relative angles which were adjusted during the direct sunlight regression.
Besides HDRF, the outputs also included the observation angles for each image and the direct sunlight ratio.
Further evaluation steps and plotting were done with Python.

\begin{figure}
  \centering
  \includegraphics[width=.48\textwidth]{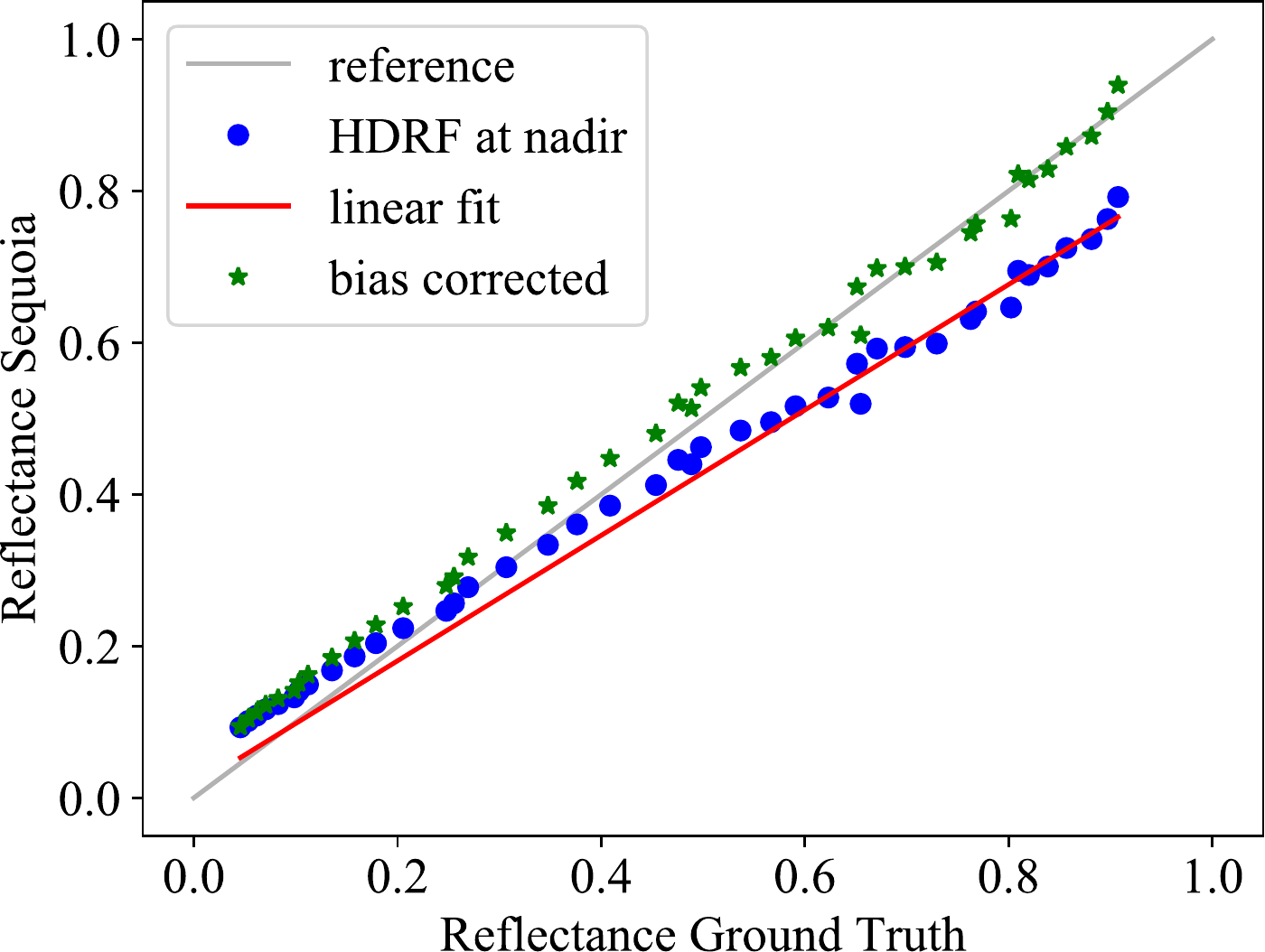}
  \caption{
    Visualisation of the bias correction procedure (here for the NIR data of the 12pm acquisition).
    A line (red line) was fitted into the uncorrected multi-grey target data using the nadir views (blue dots).
    The resulting function was applied to each data point.
    Green stars represent the nadir multi-grey target data after the bias correction.
  }
  \label{fig:bias_correction}
\end{figure}

The Sequoia camera is designed and calibrated to work with vegetation, which has typical reflectance of up to $0.5$, but not for materials with a reflectance as high as Spectralon.
In clear sky, this high reflectance brings the camera to its limits: all green frames and part of the red frames are saturated.
Due to that and the way the auto-exposure of the cameras works, we needed to manually set the exposure time of all bands to its lowest possible value (between \SI{0.3}{ms} and \SI{0.023}{ms}).
However, with such an extreme setting, the camera calibration does not seems to be valid any more, since we observed radiometric artefacts in the form of a bias in radiance (see also \cite{pix4d_wispers2018}).
Therefore we removed the bias using a separate known target composed of 42 different grey patches (6 rows by 7 columns) (``multi-grey target'') placed close to the Spectralon.
The true reflectance of each patch was measured with the spectro-radiometer viewing nadir.
Images taken nadir were selected and the HDRF was estimated for each patch.
Then a linear regression between the estimated and the true HDRF was ran.
This fit resulted in a linear transformation that was used to correct the reflectance for all Spectralon measurements.
The procedure is visualised in \cref{fig:bias_correction}.
This was possible for most of the clear sky acquisitions.

Except if mentioned otherwise, all four bands behave similarly.
Thus by default, we show detailed results only for one band and report a summary for all bands.

\section{Results and discussion}
\subsection{Direct sunlight ratio estimation}

One of the key components to obtain accurate HDRF is the direct sunlight ratio $\epsilon$.
Depending on the sky condition, the typical behaviour of the sky illumination is different.
In this study, we limit ourselves to clear sky and overcast sky.

The relative angle between the camera and the sky sensor was optimised for clear sky conditions.
For overcast sky, the angle dependency of the downwelling irradiance is very low and the available data do not constrain the angle well enough to adjust it in the fit, but small inaccuracies of the relative angles also do not have an effect.

\begin{figure}
  \begin{minipage}[b]{.48\linewidth}
    \centering
    \includegraphics[width=\linewidth]{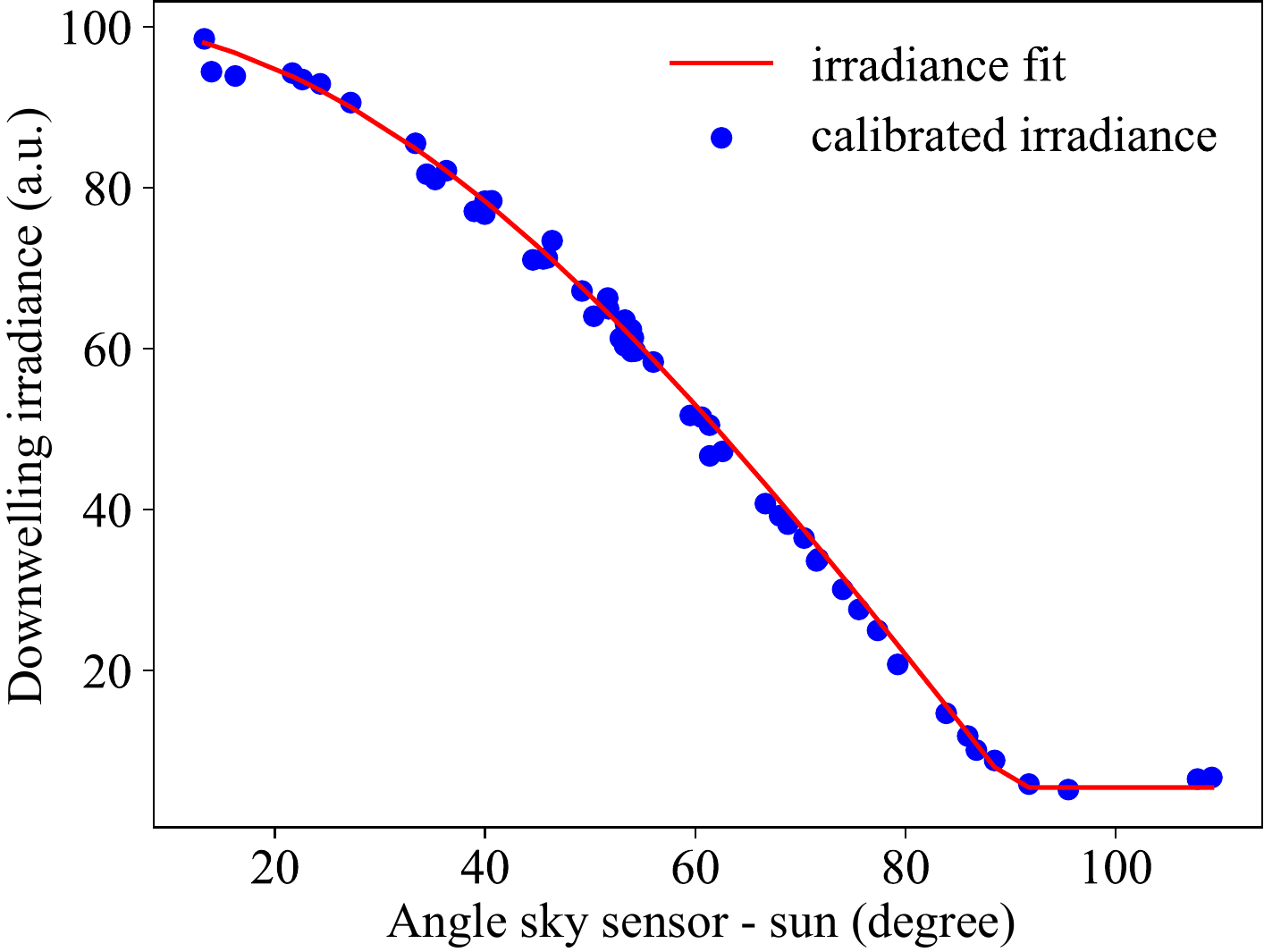}
    (a) Clear sky
    \medskip
  \end{minipage}
  \hfill
  \begin{minipage}[b]{0.48\linewidth}
    \centering
    \includegraphics[width = \linewidth]{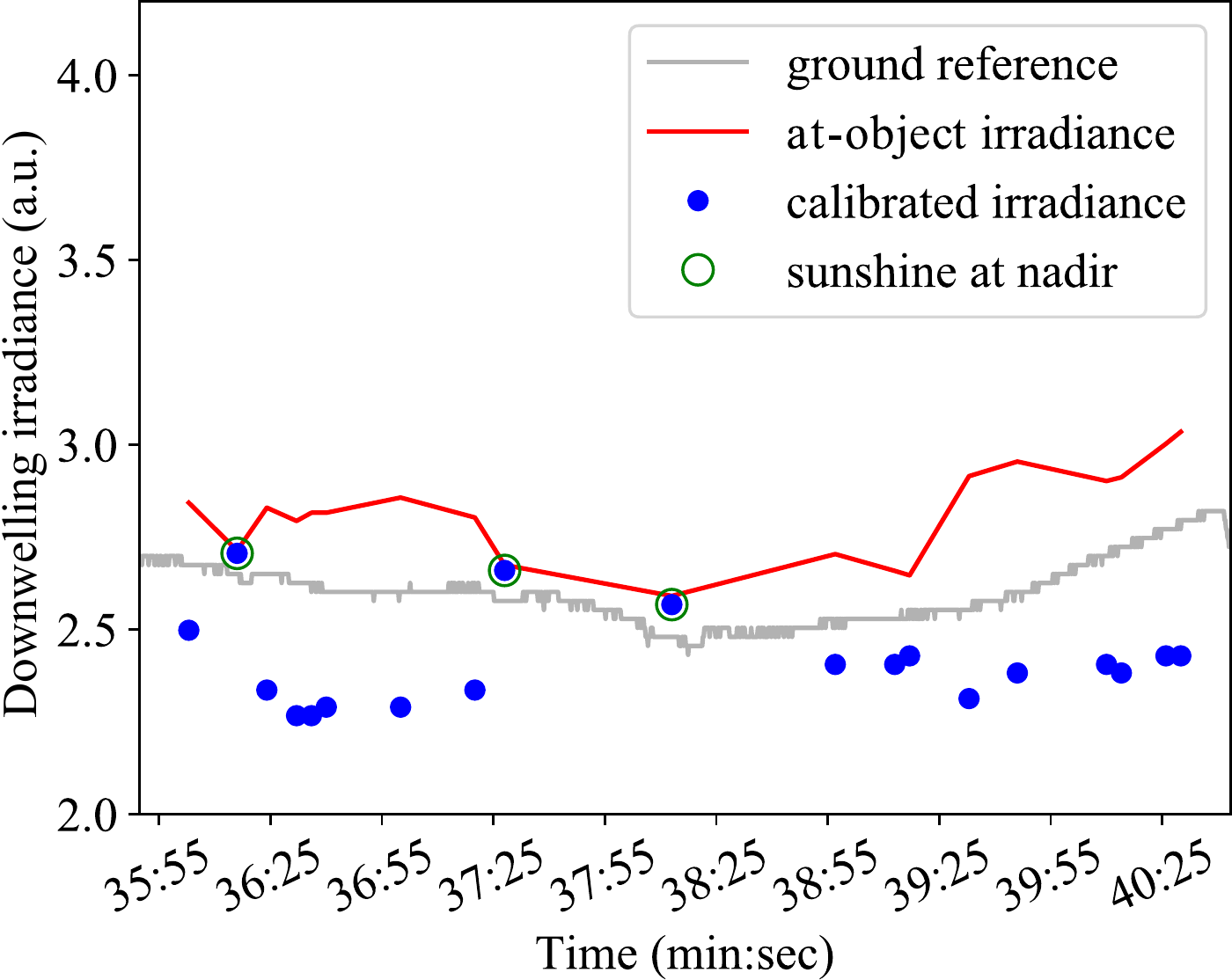}
    (b) Overcast sky
    \medskip
  \end{minipage}
  \caption{
Evolution of measured downwelling irradiance (here for the red band) with respect to the relative angle between the sky sensor and the sun for clear sky (a) and with respect to time for overcast sky (b), both in the same arbitrary irradiance units.
For clear sky (a), the irradiance fit (red line) is compared to the at-sensor irradiance measurement (blue dots).
For overcast sky (b), the at-object irradiance (i.e.\ corrected for angle effects) (red line) is compared to both the at-sensor irradiance measurement (blue dots) and the ground reference measurement (grey line).
}
  \label{fig:sun_vs_angle_regression}
\end{figure}

\Cref{fig:sun_vs_angle_regression} shows the regression for clear sky and overcast for the red band.
For clear sky, the downwelling irradiance is high and since most of the irradiance is direct sunlight, a strong relationship between measured irradiance and relative angle between the sky sensor and the sun is observed.
The fit characterised well the angle dependency of the measurement.
With angles higher than \SI{90}{\degree}, the sky sensor is only exposed to scattered light (flat line with between \SI{5}{\percent} to \SI{10}{\percent} of the total irradiance).
For short times like the acquisitions performed for this study, the total downwelling irradiance is approximately constant in time.
Thus the total downwelling irradiance retrieved from the fit can be used for the whole acquisition.

For overcast conditions, 
the downwelling irradiance is relatively low due to absorption in the clouds.
Light that is coming to the ground has been scattered many times in the clouds and therefore is nearly entirely diffuse.
As expected, the fitted direct sunlight ratio $\epsilon$ is close to $0$ ($0.06$ for Green, $0.04$ for Red and $0.07$ for Red Edge and NIR).
The total downwelling irradiance is a smoothly changing irregular function of time, because the thickness of the cloud layer changes over time.
Therefore the sky sensor correction has to be computed per frame.

We placed an additional sky sensor beside the target to compare the temporal evolution of the reconstructed at-object downwelling irradiance with a real measurement (grey line in \cref{fig:sun_vs_angle_regression}~(b)).
For most measurements, the at-sensor irradiance is below the ground reference, while nadir measurements fit nearly perfectly.
As discussed in \cref{sec:sun-model}, if the sky sensor is not pointing zenith, it measures irradiance only from a fraction of the sky, which is taken into account in our downwelling irradiance computation.
The corrected at-object downwelling irradiance (red line in \cref{fig:sun_vs_angle_regression}~(b)) matches better with the real temporal change.
Both reference and corrected irradiances decrease between 35:55 and 38:25 and increase between 38:55 and 40:25.
A slight bias is visible between the measurement and the reference that could be attributed to a limitation of our overcast model.
In real conditions, the scattered light is not uniformly distributed over the hemisphere as assumed in our model.
The implication of this misestimation for the HDRF computation is explained in \cref{sec:overcast_analysis}.

On both weather conditions, the total downwelling irradiance is well retrieved and used to obtain the HDRF of Spectralon.

\begin{figure}
	\centering
	\includegraphics[width=.70\textwidth]{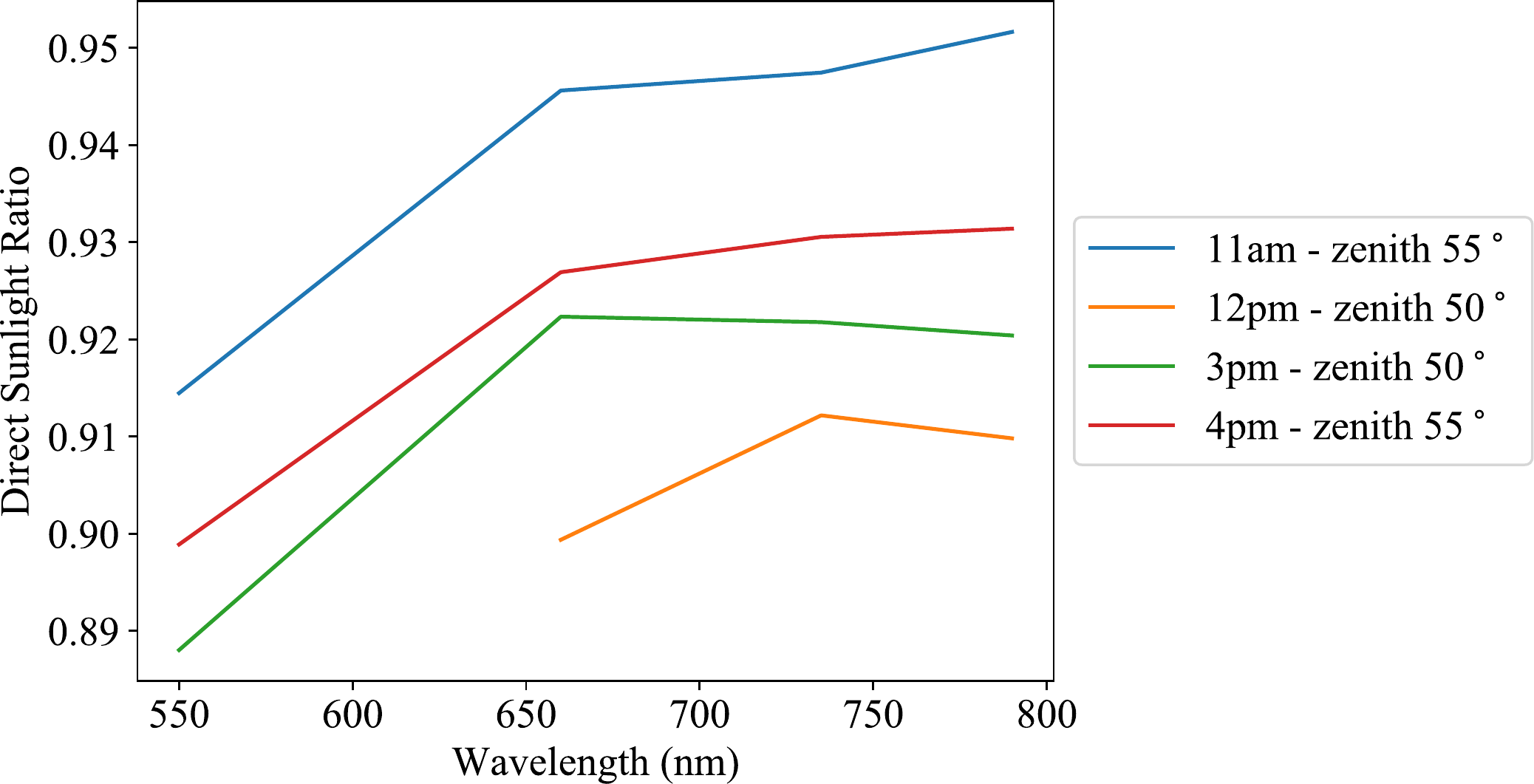}
	\caption{Evolution of direct sunlight ratio with wavelength in clear sky conditions for each of the 4 acquisitions.}
	\label{fig:directsunlightratio_vs_wavelength}
\end{figure}

In clear sky, the direct sunlight ratio is expected to vary with wavelength: blue light is scattered more than red light (it is dominated by Rayleigh scattering, whose cross section is proportional $\lambda^{-4}$, which is why the sky is blue).
The estimated direct sunlight ratio as a function of wavelength is shown in \cref{fig:directsunlightratio_vs_wavelength}.
The value for the green band at 12pm is not shown because the sun angle fitting algorithm did not converge for this particular measurement.
A clear wavelength dependency with increasing values for increasing wavelength is observed, which corresponds well to the expected behaviour.

No clear relationship of the direct sunlight ratio with respect to the sun zenith angle is observed, as the range of acquired sun zenith angle is too small to notice such relationship.

\subsection{Spectralon HDRF estimation}

Three metrics were computed to assess the quality of the multi-angular HDRF estimation: the Pearson correlation coefficient (PCC), the Root Mean Square Error
\begin{equation}
\text{RMSE} = \sqrt{\frac{1}{N} \sum_{i=1}^{N}E_i^2}
\end{equation}
and the Standard Deviation Error
\begin{equation}
\text{SDE} =\sqrt{\frac{1}{N-1}\sum_{i=1}^{N}(E_i-\overline{E})^2 }
\,,
\end{equation}
where the error $E_i$ is defined as the difference in HDRF between the result of our retrieval method and the literature value (HDRF from the database for clear sky, constant across all viewing angles for overcast) and $\overline{E}$ denotes the mean error.
The RMSE is computed to estimate the global accuracy of the computation.
The SDE is a measure for the spread of the error for the different data points and is used to assess the validity of the method: if the SDE is in the same order of magnitude as the RMSE, the error is random. If it is much lower, there is an approximately a constant offset. 

Different Spectralon targets have slightly different characteristics.
We quantified the error introduced by the difference between our Spectralon and the Spectralon used by \citet{bruegge+2001} by comparing the directional-hemispherical reflectance factor (DHRF) with \SI{8}{\degree} incident angle.
The value at \SI{632.8}{nm} for the unit used by \citet{bruegge+2001} is equal to $0.991$ (as measured with a laser).
The respective value for our Spectralon (derived from the DHRF provided with the calibration certificate from Labsphere) is equal to $0.9896$, resulting in a difference of around \SI{0.1}{\%}.
This is way beyond the measurement uncertainty of our setup and can be neglected.

We also investigated the spectral variation of the four different bands.
From the \SI{8}{\degree} DHRF data provided by Labsphere, the spectral variation is around \SI{0.1}{\percent}.
\citet{chrien+1998} measured the spectral difference in BRF for incident angles of \SI{55}{\degree} and \SI{50}{\degree} and found differences of up to \SI{2.5}{\percent} for high viewing angles (\SI{50}{\degree}), which is equal to the measurement accuracy of the cameras.
This last error is the main source of difference and explains a possible bias of computed reflectance between different bands.

\subsubsection{Clear sky data analysis}

For the clear sky data acquisition, the HDRF of Spectralon is computed using $E_\text{sun}$, the total downwelling irradiance projected onto the object of interest (cf.\ \cref{eq:obj-sun-irradiance}).
It is then compared with the BRF database.
We use the data with an incident angle of \SI{55}{\degree} for the acquisitions at 11am and 4pm (solar zenith angle at \SI{53}{\degree} and \SI{55}{\degree}, respectively) and an incident angle of \SI{50}{\degree} for the acquisitions at 12pm and 3pm (solar zenith angle at \SI{49}{\degree} and \SI{51}{\degree}, respectively).

The three metrics used to characterise the quality of the multi-angular HDRF retrieval are summarised in \cref{tab:clearsky_results} for all acquisitions except the 11am acquisition where the multi-grey target was not available in the frames.
For this particular acquisition, only PCC and SDE were computed.

\begin{figure}
  \begin{minipage}[b]{.48\linewidth}
    \centering
    \includegraphics[width = \linewidth]{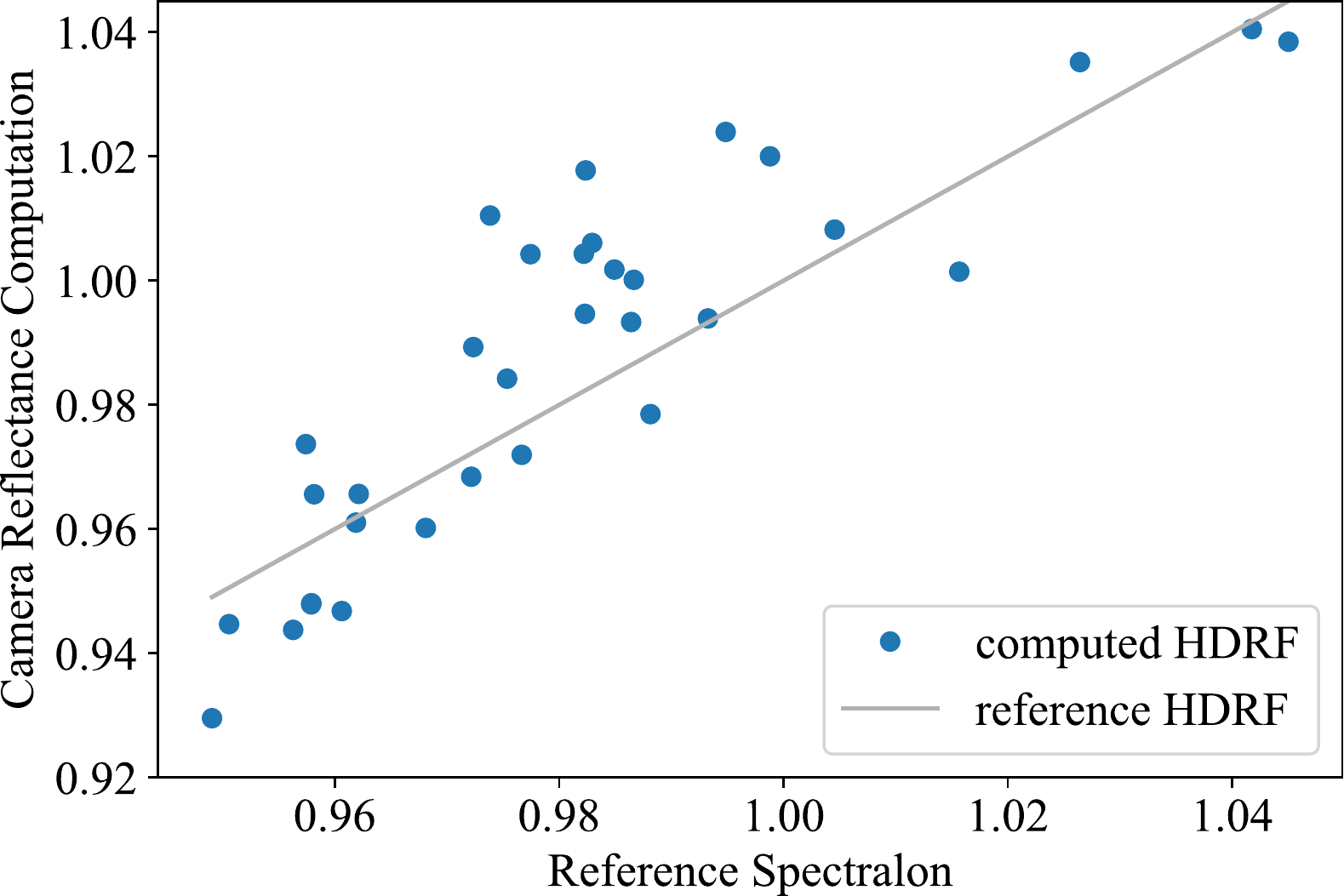}
    (a) Comparison with the literature values
  \end{minipage}
  \hfill
  \begin{minipage}[b]{0.48\linewidth}
    \centering
    \includegraphics[width = \linewidth]{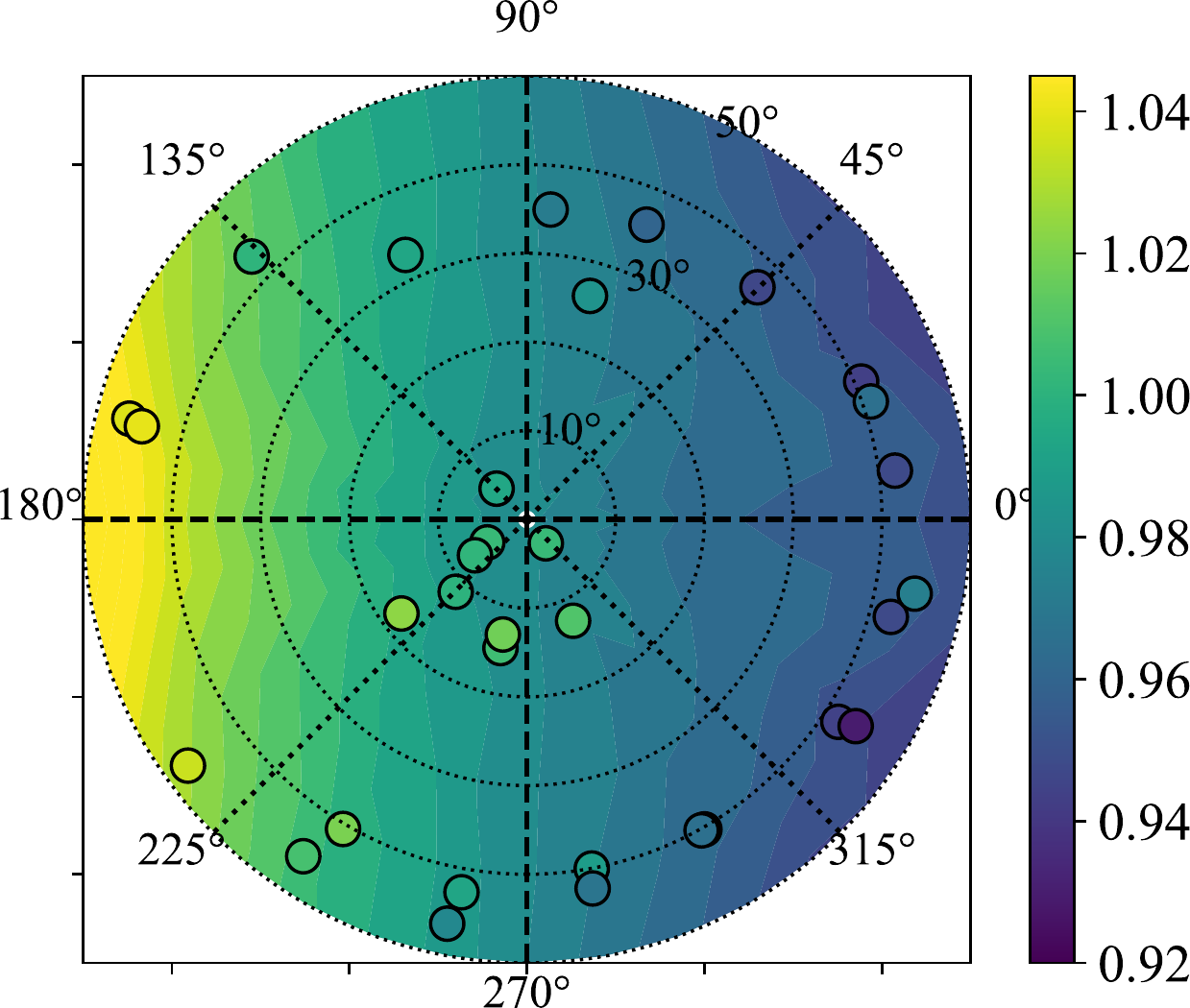}
    (b) HDRF as a function of the viewing angle
  \end{minipage}
  \caption{
    Comparison of the estimated HDRF of Spectralon with the laboratory measurement by \citet{bruegge+2001} for the NIR band of the measurement at 12pm. 
    (a) true vs.\ the estimated values with the ideal 1:1 case shown as grey line.
    (b) Visualisation of the HDRF as a function of the viewing angle in solar coordinates, with the literature BRF values in the background and the values retrieved by our measurement as coloured dots.
    The concentric circles and dotted lines represent the $\theta$ and $\varphi$ viewing angles, respectively ($\theta$ between \SI{0}{\degree} and \SI{50}{\degree}, and $\varphi$ between \SI{0}{\degree} and \SI{360}{\degree}).}
  \label{fig:clearsky_hdrf}
\end{figure}

\begin{table}
  \caption{The Root Mean Square Error (RMSE) and the Standard Deviation Error (SDE) between the estimated and true HDRF for the clear sky acquisitions.}
  \centering
  \begin{tabular}{ l l c c c }
      \hline
      Time - Solar Zenith Angle & Band & RMSE (\%) & SDE (\%) & PCC \\ \hline
      \multirow{3}{*}{11am - \SI{53}{\degree}} & Red & - & 1.29 & 0.91\\
      & Red edge & - & 1.32 & 0.92\\
      & NIR & - & 1.51 & 0.90\\ \hline
      \multirow{3}{*}{12pm - \SI{49}{\degree}} & Red & 1.79 & 1.58 & 0.89\\
      & Red edge & 1.60 & 1.18 & 0.91\\
      & NIR & 1.60 & 1.52 & 0.86\\ \hline
      \multirow{3}{*}{3pm - \SI{51}{\degree}} & Red & 6.04 & 1.04 & 0.94\\
      & Red edge & 4.42 & 0.97 & 0.97\\
      & NIR & 5.39 & 1.26 & 0.93\\ \hline
      \multirow{3}{*}{4pm - \SI{55}{\degree}} & Red & 1.61& 1.62&0.81\\
      & Red edge &2.05&1.42&0.87\\
      & NIR & 3.20& 1.69&0.83\\ \hline
  \end{tabular}
  \label{tab:clearsky_results}
\end{table}

A strong correlation between the reference and the computed HDRF was found,
with a correlation coefficient between $0.81$ and $0.94$ for Red, between $0.87$ and $0.97$ for Red Edge and between $0.83$ and $0.93$ for NIR.
This shows that the bi-directionality of the Spectralon is well retrieved.
The RMSE is varying per acquisition but is quite constant per band when considering a single acquisition.
It is between \SI{1.6}{\percent} and \SI{6.1}{\percent} which corresponds to the errors both of the model and the camera.
Even for the measurement with largest bias (the one at 3pm), the HDRF estimation is strongly correlated with the true value,
with a correlation coefficient between $0.94$ and $0.97$.
The SDE is ranging between \SI{1.0}{\percent} and \SI{1.7}{\percent}. These values will be used as reference for the overcast acquisition analysis.

The band that correlates best with the laboratory measurement is red edge, for most of the acquisitions.
This might be due to the small bandwidth that allowed us to acquire the data at a higher exposure time, where the camera performs better.

A visualisation of the estimated HDRF with respect to the literature laboratory measurement is presented in \cref{fig:clearsky_hdrf} for the NIR band of the 12pm acquisition.
$\varphi = \SI{180}{\degree}$ and $\theta = \SI{50}{\degree}$ represent the direct solar reflection.
As discussed above, the bi-directionality of the reflectance is well recovered and matches the laboratory measurements.

\subsubsection{Overcast sky data analysis}
\label{sec:overcast_analysis}

For the overcast sky acquisition, the HDRF of Spectralon is expected to be constant for all viewing angles.
That is characteristic for an almost Lambertian surface like Spectralon.

In addition to the setup for the clear sky acquisition, an additional sky sensor is located close to the target to measure the at-target downwelling irradiance.
This allows to compute an HDRF using the measured at-target downwelling irradiance instead of using the at-sensor downwelling irradiance and back-computing the at-target downwelling irradiance.
While this is not a typical use case for UAV acquisitions, it allows us to better evaluate the influence of the sky model and the angle correction on the HDRF estimation.

\begin{figure}
  \begin{minipage}[b]{.48\linewidth}
    \centering
    \includegraphics[width = \linewidth]{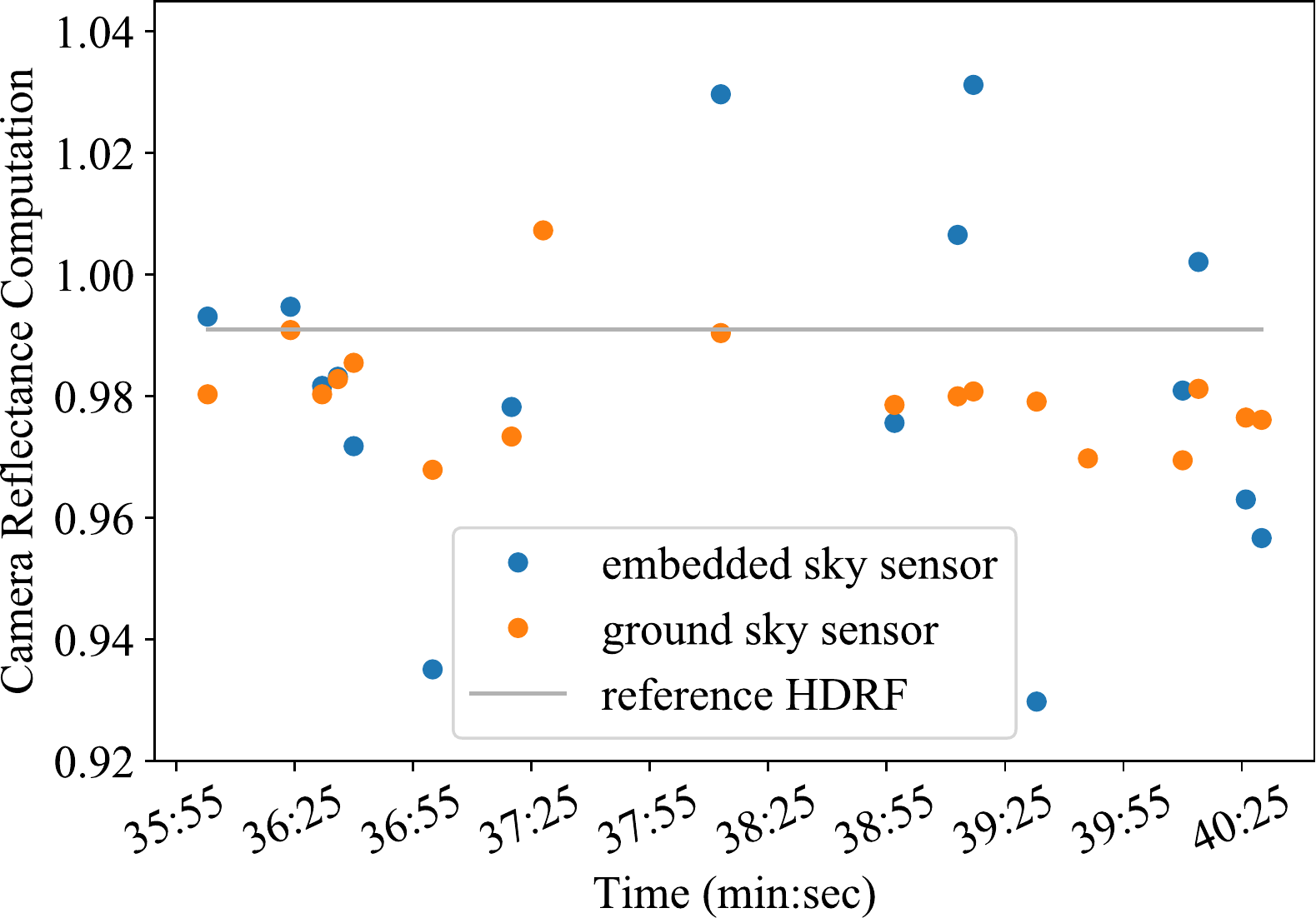}
    (a) Estimated HDRF as a function of time
  \end{minipage}
  \hfill
  \begin{minipage}[b]{0.48\linewidth}
    \centering
    \includegraphics[width = \linewidth]{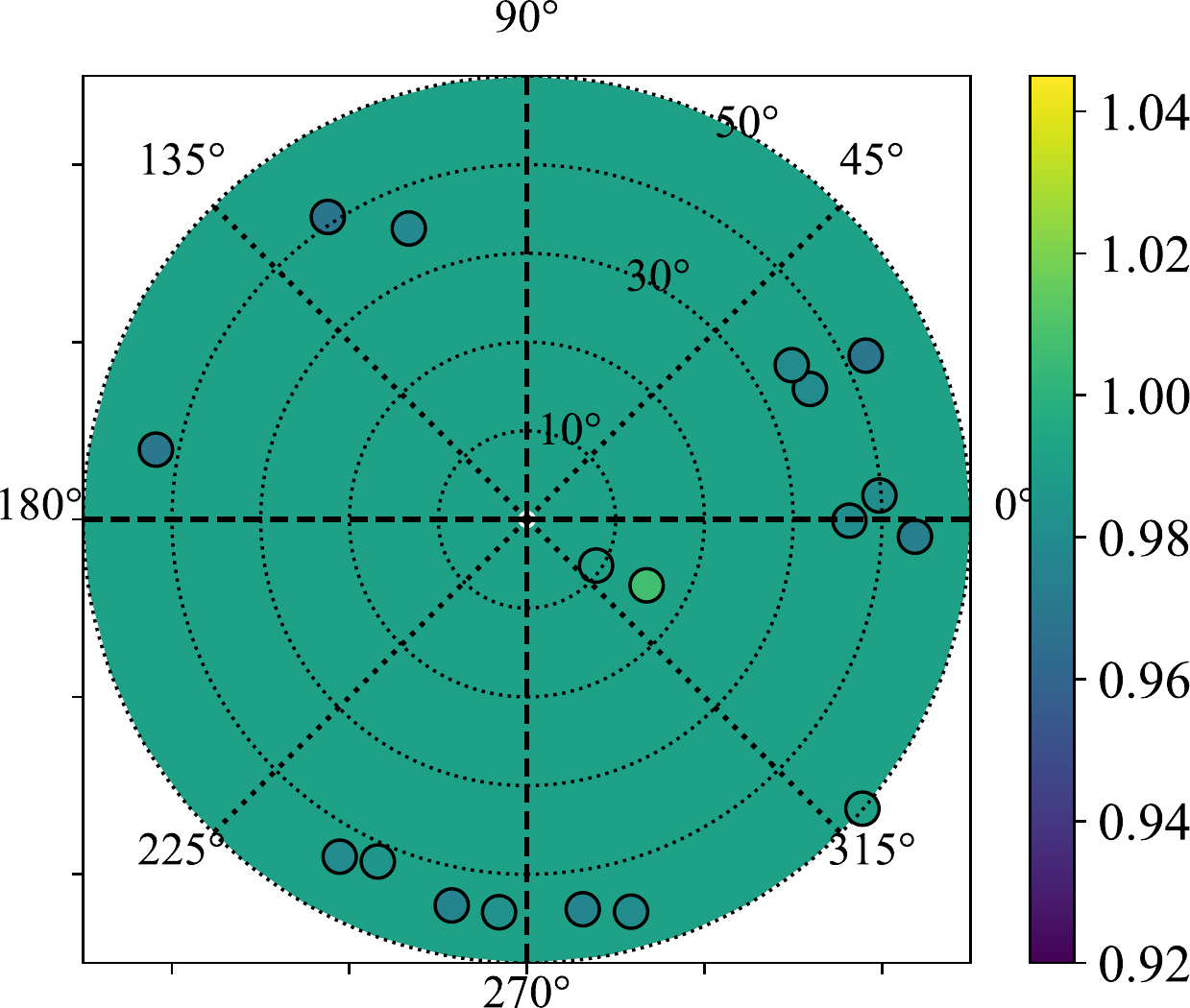}
    (b) HDRF as a function of the viewing angle
  \end{minipage}
  \caption{
    Comparison of the computed HDRF of Spectralon with the laboratory measurement as reference. 
    (a) Estimated HDRF as a function of time, with blue dots showing values retrieved using the sky sensor fixed on the camera and orange dots using the ground sky sensor.
    (b) Visualisation of the HDRF as a function of viewing angles, with the reference value in the background and the retrieved values using the ground sunshine sensor as coloured dots.}
  \label{fig:overcast_hdrf}
\end{figure}

\begin{table}
  \caption{The Root Mean Square Error (RMSE) and the Standard Deviation Error (SDE) for the overcast acquisition, both for using the sky sensor fixed to the camera and for using the ground sky sensor to compute the HDRF.}
  \centering
  \begin{tabular}{@{\extracolsep{4pt}} l c c c c @{}}
      \hline
	& \multicolumn{2}{c}{embedded sky sensor} & \multicolumn{2}{c}{ground sky sensor} \\
	\cline{2-3} \cline{4-5}
      Band & RMSE (\%) & SDE (\%) & RMSE (\%) & SDE (\%)\\ \hline
      Green & 3.65 & 3.63 & 1.37 & 0.92 \\
      Red & 4.54 & 3.36 & 3.46 & 1.20 \\
      Red edge & 4.25 & 3.63 & 2.46 & 1.05 \\
      NIR & 5.04 & 3.67 & 3.59 & 1.14 \\ \hline
  \end{tabular}
  \label{tab:overcast_results}
\end{table}

\Cref{fig:overcast_hdrf} shows the estimated HDRF for both approaches. 
The RMSE and SDE for both evaluations is shown in \cref{tab:overcast_results}.
The PCC could not be computed because the true HDRF is constant across the viewing angles.

The true HDRF is well retrieved, with RMSE ranging from \SI{1.3}{\percent} to \SI{5.0}{\percent}.
Looking at the SDE, the at-ground measurement yields better results, decreasing SDE from \SI{3.5}{\percent} for using the sky sensor fixed on the camera to \SI{1.0}{\percent} for using the ground sky sensor, which is closer to the SDE computed for clear sky.
This is attributed to the assumption of a uniform diffuse illumination.
Due to moving clouds with different thickness, the real illumination is irregular.
If the sky sensor is oriented nadir and measures the whole hemisphere, the effective downwelling light is well retrieved.
Once the sky sensor is orientated at high angles, a certain portion of the sky is not seen and its contribution to the total downwelling irradiance must be estimated using the value measured in the visible part of the sky.
However the real angular distribution of the downwelling irradiance is difficult to model without measuring the cloud distribution (for example by mapping the clouds).
This is expected to be less problematic for real flight datasets, since images are typically acquired more or less nadir.

\section{Conclusions}

We presented a new physically based method to estimate the HDRF from images of lightweight multispectral cameras with a downwelling irradiance sensor aboard UAVs which combines radiometry with photogrammetric computer vision to derive geometrically and radiometrically accurate data purely from the images. It does not require calibration targets, works both in clear sky and overcast conditions, and allows to capture the directionality of the reflectance factor.
The method works for any calibrated multispectral camera with a sky sensor that is rigidly connected to the camera.
It was tested with a ground acquisition of a well-characterised Spectralon target.
We used the Parrot Sequoia with the downwelling irradiance sensor fixed opposite the camera, a configuration typically found on UAVs, and acquired data with different orientation all around the target to obtain different viewing angles in clear sky and overcast conditions.
A good agreement of the HDRF estimated with our method for the different viewing directions with literature values was found, with an RMSE of \SI{3.0}{\percent} for clear sky and \SI{2.7}{\percent} for overcast sky.
For overcast sky, the assumption of isotropic scattered light was found to lead to a small additional error for non-nadir measurements, which increases the error to a maximum of \SI{5}{\percent}.
Our measurement shows that our method is capable of retrieving correct HDRF in various conditions.
Future research should check the performance for real flight data and study the dependency of heterogeneity of scene materials or terrains slope in the retrieved HDRF.

\section*{References}
\bibliographystyle{elsarticle-harv}

\end{document}